# High-Field Conduction in Barium Titanate

F. D. Morrison, P. Zubko, D. J. Jung, and J. F. Scott[*]

Centre for Ferroics, Earth Sciences Department, University of Cambridge,

Cambridge CB2 3EQ, U. K.

P. Baxter, M. M. Saad, J. M. Gregg and R. M. Bowman,

Physics Dept., Queen's University, Belfast BT7 1NN, U. K.

**Abstract**

We present current-voltage studies of very thin (*ca.* 77 nm) barium titanate single crystals up to 1.3 GV/m applied field. These show that the mechanism of leakage current at high fields is that of space charge limited conduction (SCLC) in a regime with a continuous distribution of traps, according to the original model of Rose [Phys. Rev. 97, 1537 (1955)]. This study represents a factor of x5 in field compared with the early studies of $BaTiO_3$ conduction [A. Branwood et al., Proc. Phys. Soc. London 79, 1161 (1962)]. Comparison is also given with ceramic multilayer barium titanate capacitors, which show similar SCLC behaviour. The data are shown to be completely incompatible with variable range hopping [B. I. Shklovskii, Sov. Phys. Semicond. 6, 1964 (1973)], despite the recent report of such mechanism in $SrTiO_3$ films [D. Fuchs, M. Adam, and R. Schneider, J. Phys. IV France 11, 71 (2001)].

---

[*] email: jsco99@esc.cam.ac.uk

The very recent preparation of ultra-thin (77 ± 10 nm) single crystals of barium titanate via focussed-ion-beam techniques [1] has permitted studies of the dielectric properties and phase transitions in thin ferroelectric oxides with no substrate or related strain and perfectly symmetric top and bottom electrodes (simultaneously sputtered).  Such a system is a text-book example of a ferroelectric capacitor without extrinsic complications such as asymmetric electrodes, large near-electrode gradients of strain and/or oxygen vacancies, or "dead" layers of chemically distinct interfacial material.  Indeed, the dielectric behavior show these specimens to exhibit bulk properties without the usual broadening or shift in transition temperature.

Such samples permit a re-examination also of the conduction properties (leakage currents) in classic ferroelectric oxides.  The classic work on barium titanate [2] exemplified in the Landolt-Bornstein Encyclopedia of Physics shows studies with different top and bottom electrode combinations but was limited to 200 MV/m.  High-field behavior in thin-film perovskite oxides is at present controversial, with Fuchs et al. [3] claiming variable-range hopping mechanisms in strontium titanate.  According to this model [4-6], the current-voltage behavior is given by

$$I(E) = A \exp[-(E^*/E)^{1/4}], \qquad (1)$$

where A is a temperature dependent coefficient and E* a critical field characteristic of the material and possibly its electrodes.

Typical results from the present work are shown in Fig.1a.  Here an external field of up to 100 V is applied over 77 nm, producing a nominal maximum field of 1.3 GV/m.  The data appear to fall into three regimes:  Above 200 MV/m the behavior satisfies a space charge limited current model, where the voltage exponent is very nearly 4.0.  As initially shown by Rose [7], this is the behavior expected for a continuous distribution of traps. Such regimes must have voltage exponents between 3.0 and 4.0, as shown by Lampert and Mark [8] and in Scott's text [9].

Even when current-voltage behavior is space charge limited, data usually do not agree with the simplest quadratic voltage dependence predicted from the Mott-Gurney Law. In general, several corrections to the Mott-Gurney Law are known to be applicable: Rather than a simple $I(V) = a V^2/d^3$ dependence, where V is applied voltage and d is film thickness, one instead expects more precisely that [8]

$$I(V,T) = A\, V^{(L+1)(1-ma/d)}\, (d-2a)^{-(2L+1)} \qquad (2)$$

Where L is a value near between 1 and 2; a is the accommodation length (typically 10% of d [10]); and m is a constant of order 2-3.

At intermediate fields the current-voltage behavior is linear, with exponent very nearly 1.0; this is not ohmic and is the behavior predicted by Simmons [11] for Schottky-limited current in systems where the mean free path of electrons injected from the cathode is less than the width of the Schottky barrier (the opposite approximation was made in the original Schottky model and the resulting equation is often unwisely used to describe conduction in oxides, yielding unphysical Richardson coefficients and misinterpreted "ohmic" conduction).

$$J = \alpha\, T^{3/2}\, E\, \mu\, (m^*/m)\, \exp[-(\Phi/kT)\, \beta\, E^{1/2}], \qquad (3)$$

where $\mu$ is electron mobility; $m^*$, effective mass; $\Phi$, work function and E, field across the dielectric (not necessarily the same as total applied field V/d).

At very low voltages there are apparently some trap states, producing deviations from linearity as originally shown by Hamann et al. [12] and reproduced in Scott's text [13]. Fig. 1a illustrates the low-field region with trap discontinuities; a comparison is shown with the model of Hamann et al, figure 1b. This comparison is only qualitative and pedagogic, but it shows what is expected: regimes of linear and quadratic dependences, separated by either changes in slope or (smoothed) step discontinuities.

We note that the applied field of 1.3 GV/m actually exceeds the theoretical estimate of breakdown field $E_B$ for such perovskite oxide films of 800 MV/m given by Scott [14]:

$$E_B(T, t_c) = [3C_V K/\sigma_0 \beta t_c]^{1/2} T \exp [\beta/2kT], \qquad (4)$$

Where the breakdown field E is a function of temperature T and voltage ramp time $t_c$, and the other parameters have their usual meanings: $C_V$ is specific heat; K, thermal conductivity; k, Boltzmann's constant; and $\sigma_0$ and $\beta$ defined by the exponential conductivity in the non-ohmic regime

$$\sigma(T) = \sigma_0 \exp [-\beta/kT]. \qquad (5)$$

This equation predicts breakdown near E(internal) = 800 MV/m in $BaTiO_3$ or $Ba_{1-x}Sr_xTiO_3$ films, using empirical parameters for the constants in Eqs.3,4. The fact that present data exceed 1.3 GV/m for the E(external) applied is probably due to the fact that for a 77 nm film of dielectric constant $\varepsilon$ = 2700, approx. 50% of the field will drop across the Au electrodes, and only 50% across the dielectric [15]. Therefore the present study reaches only ca. 650 MV/m internal field, still slightly below the estimated breakdown value of 800 MV/m.

As a comparison with the single-crystal data above, we have measured some ceramic barium titanate multilayer ceramic films of 10 ± 1 microns thickness from AVX Corporation. The details of those materials will be presented elsewhere, but Fig. 2a illustrates the plateau that occurs over an extended region of applied voltage, with Fig.2b illustrating a fit to Dawber's diffusion model at low fields (<60 V; <6 MV/m) and Simmons' model at high fields (>60 V; ; >6 MV/m). These data were obtained with a 1.0 s soak time; other data at longer or shorter times show that the system is not in true steady state, such that fast electronic processes are measured but not slow ionic transport or slow trap filling/emptying. Fig.3a demonstrates that Dawber's diffusion equation [16] below gives a very good fit to the low-field data; Dawber's model does not include slow ionic diffusion:

$$I(V) = \mu N E \exp[-(q/kT) \phi_B - (qN_D^+ w^2/2\varepsilon) + (\varepsilon/2qN_D^+)E^2], \qquad (6)$$

where $\mu$ is the electron mobility; N, the number of carriers of charge q; $N_D^+$, the number of ionized donors, which is a function of temperature; $\phi_B$, the Schottky barrier height; w, the depletion width; $\varepsilon$ the d.c. dielectric constant; and E, the field in the interior of the film (modified from V/d by screening, polarization, and voltage drops in the electrodes and interfaces). Note that this equation predicts, with the usual approximation $E \cong V/d$, that $\log(I/V) = \text{const.} \times V^2$, a relationship we show experimentally below.

Fig.3b shows that up to 100 V their behavior does not satisfy variable-range hopping models; a straight line would be obtained from Eq.1 if the variable-range hopping model were correct. These data are in the same field regime as those of Fuchs et al. on strontium titanate [3] but in contrast with the data of those authors do not satisfy the variable-range hopping model.[4-6]

In summary, new work on barium titanate single crystals and ceramics confirms the original notion [17-19] that space charge limited currents are dominant in perovskite oxides at high fields, a conclusion that has been occasionally controversial. The diffusion model of Dawber and Scott [16] fits the plateaux observed in I(V) data in ceramic films of $BaTiO_3$.

**Acknowledgments:**

The authors would like to thank AVX Ltd, Coleraine, Northern Ireland for supplying commercial samples.

Figure captions:

Figure 1: (a) Current-voltage relationship in 77 nm barium titanate single crystal; (b) comparison with trap model, after C. Hamann, Burghardt and Frauenheim [12]. In contrast to the linear and quadratic regimes shown in this figure, we note that H. L. Stadler and P. J. Zachminidis, J. Appl. Phys. **34**, 3255 (1963) measure $BaTiO_3$ up to 45 MV/m and find I(E) varying as $E^{-1.4}$ in the intermediate field region.

Figure 2: (a) I(V) for three different commercial $BaTiO_3$ multilayer capacitor samples. (b) Reverse-bias data for sample A (Ag-Pd electrodes) fitted to diffusion model [19] at low field (<60 V) and Simmons' equation [11] at high field (>60 V).

Figure 3: (a) Plot of log (I/V) vs $V^2$ for low voltages is linear as expected. [19] (b) The log I vs $V^{-1/4}$ plot is clearly not linear, thus ruling out the possibility of variable-range hopping model of Shklovskii at high field.

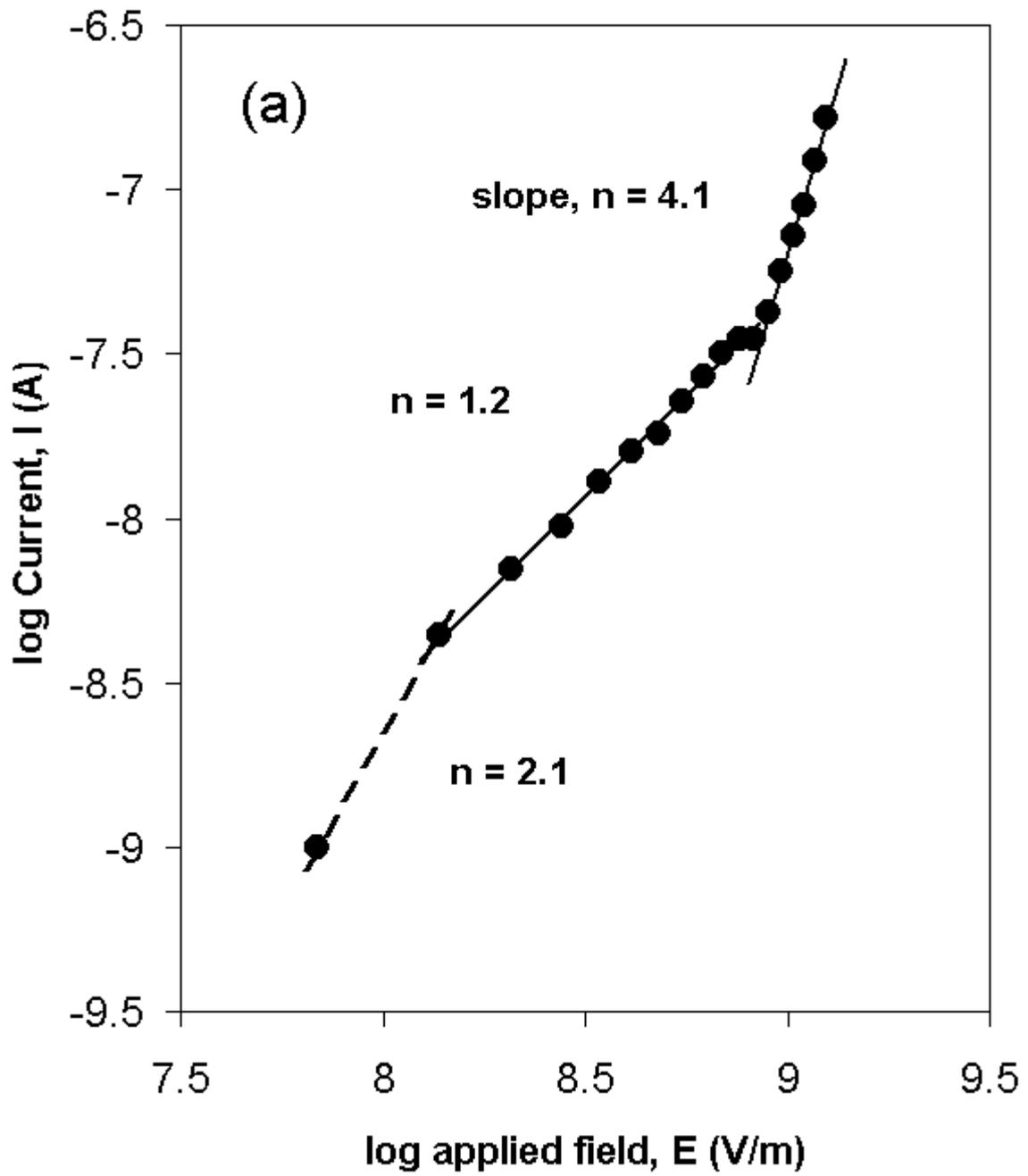

Fig 1a

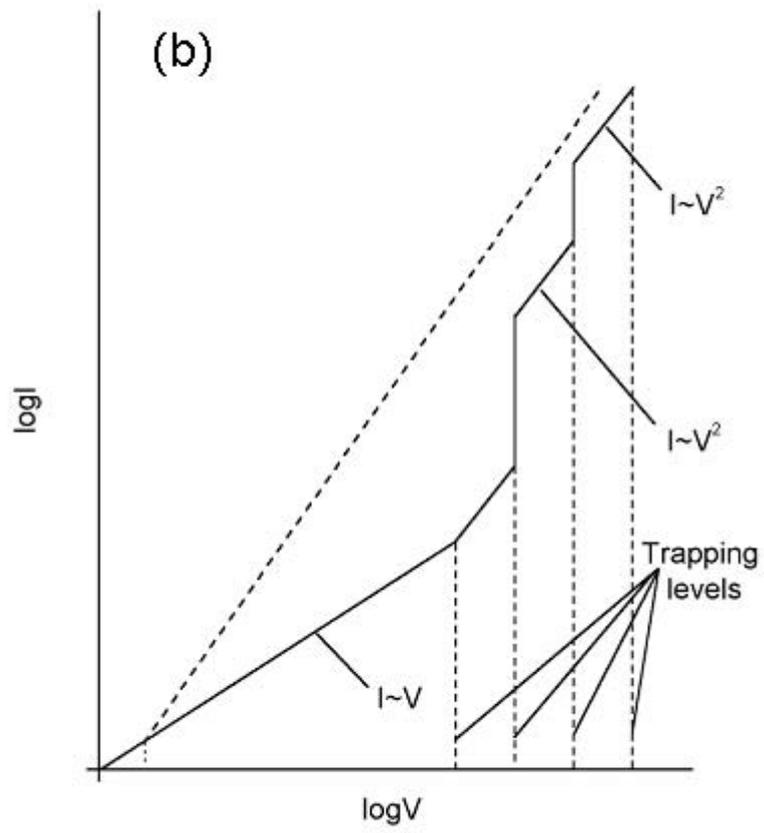

Fig 1b

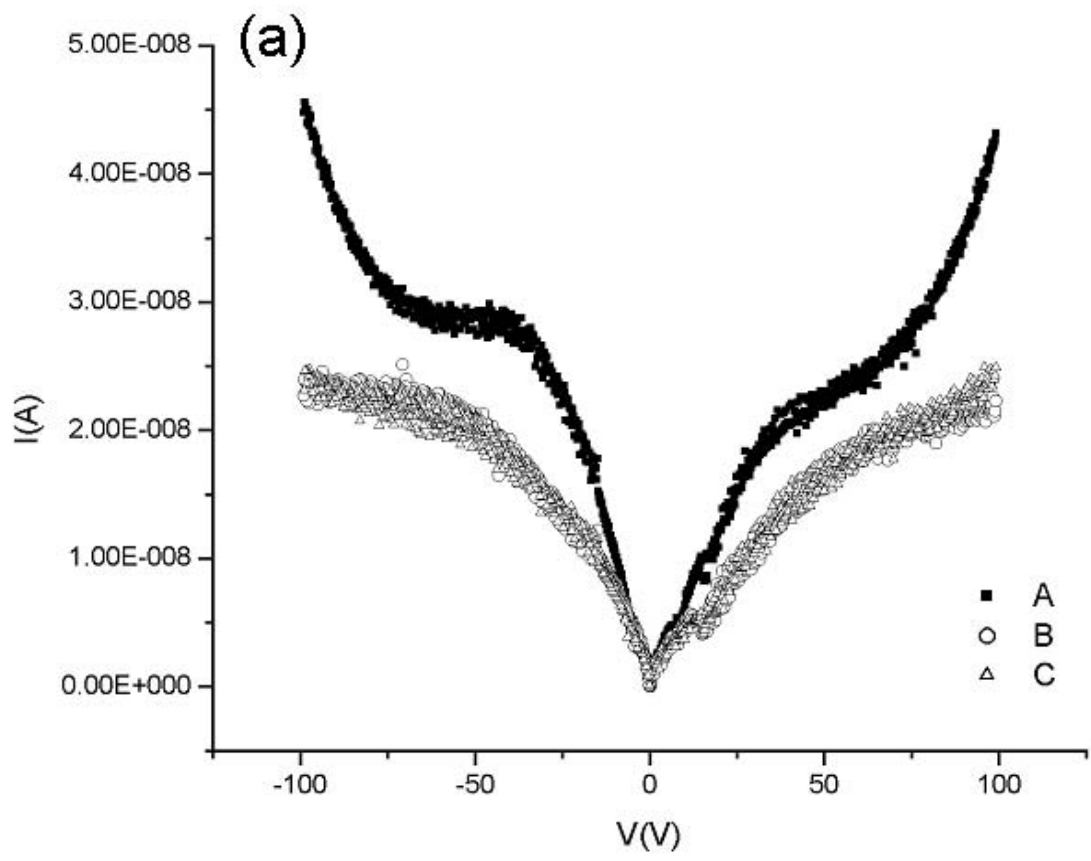

Fig2a

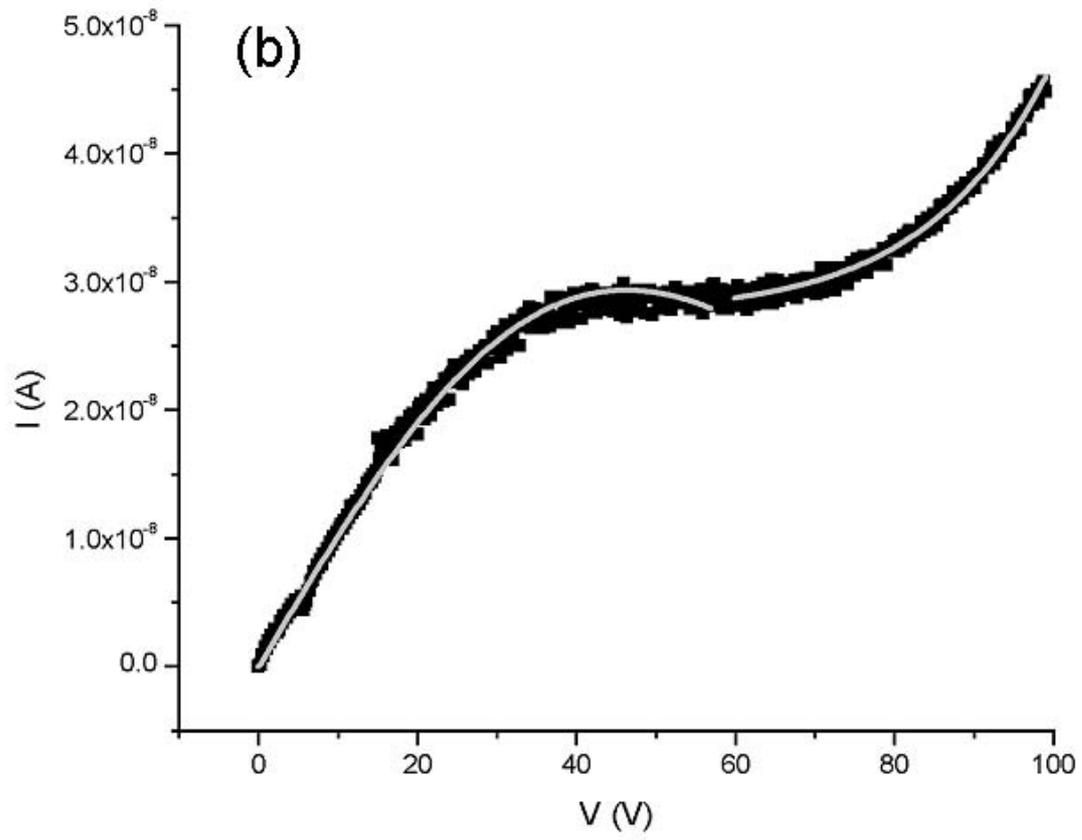

Fig 2b

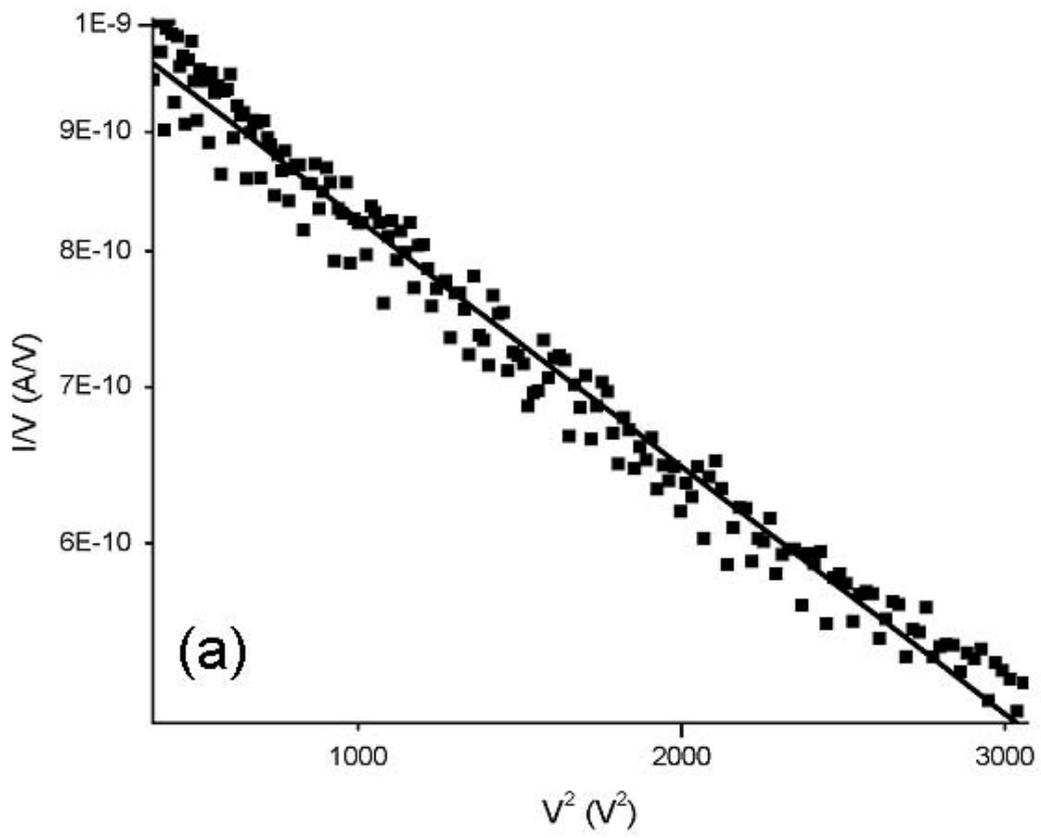

Fig 3a

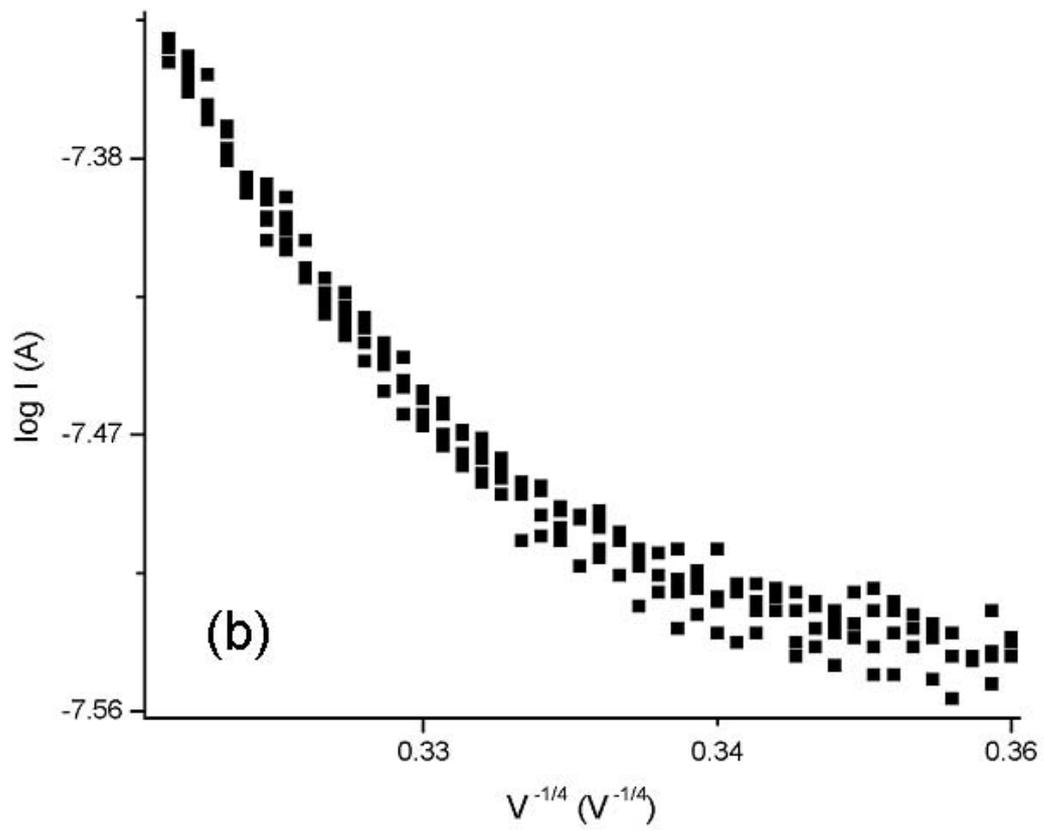

Fig 3b